\documentclass{article}
\usepackage{nips14submit_e,times}
\usepackage{amsmath, amssymb, bm, cite, epsfig, psfrag}
\usepackage{algorithm,algorithmic}
\usepackage{ifthen}
\usepackage{caption}
\usepackage{bbm}
\usepackage{subcaption,sidecap}
\usepackage{etoolbox}
\usepackage{tikz}
\usetikzlibrary{shapes,arrows,positioning,calc}

\usepackage{color}
\usepackage[normalem]{ulem} 

\newtoggle{conference}
\toggletrue{conference} 
\graphicspath{{figures/}}

\def\beq{\begin{equation}}
\def\eeq{\end{equation}}
\def\beqa{\begin{eqnarray}}
\def\eeqa{\end{eqnarray}}
\def\beqan{\begin{eqnarray*}}
\def\eeqan{\end{eqnarray*}}

\def\argmin{\mathop{\mathrm{arg\,min}}}
\def\argmax{\mathop{\mathrm{arg\,max}}}

\def\bhat{\widehat{b}}

\def\qhat{\widehat{q}}

\def\taubar{\overline{\tau}}

\def\la{\leftarrow}
\def\ra{\rightarrow}

\def\Exp{\mathbb{E}}

\def\Tm1{T\! - \! 1}
\def\Tp1{T\! + \! 1}
\def\tm1{t\! - \! 1}
\def\tp1{t\! + \! 1}
\def\km1{k\! - \! 1}
\def\kp1{k\! + \! 1}

\def\pbf{\mathbf{p}}

\def\qbf{\mathbf{q}}
\def\qbfhat{\widehat{\mathbf{q}}}
\def\rbf{\mathbf{r}}

\def\sbf{\mathbf{s}}
\def\sbfhat{\widehat{\mathbf{s}}}
\def\ubf{\mathbf{u}}

\def\vbf{\mathbf{v}}
\def\vtilde{\tilde{v}}

\def\wbf{\mathbf{w}}
\def\wbfhat{\widehat{\mathbf{w}}}
\def\xbf{\mathbf{x}}

\def\ybf{\mathbf{y}}
\def\zbf{\mathbf{z}}

\def\Wbf{\mathbf{W}}
\def\What{\widehat{W}}

\def\betabf{{\boldsymbol \beta}}

\def\taubf{{\boldsymbol \tau}}

\def\thetahat{{\widehat{\theta}}}

\newcommand{\indic}[1]{\mathbbm{1}_{ \{ {#1} \} }}

\def\Ca2{[\mbox{Ca}^{2+}]}

\title{Scalable Inference for Neuronal Connectivity from Calcium Imaging}

\author{
Alyson K.~Fletcher \\
Dept.\ Electrical Engineering \\
University of California, Santa Cruz\\
\texttt{alyson@ucsc.edu} \\
\And
Sundeep Rangan \\
Dept.\ Electrical \& Computer Engineering \\
New York University Polytechnic Institute \\
\texttt{srangan@nyu.edu}
}

\nipsfinalcopy 

\begin{document}


\maketitle

\begin{abstract}
Fluorescent calcium imaging provides a potentially powerful tool
for inferring connectivity in neural circuits with
up to thousands of neurons.
However, a key challenge in using calcium imaging for connectivity detection
is that current systems often have a temporal response and frame rate
that can be orders of magnitude slower than the underlying neural spiking
process.
Bayesian inference methods based on expectation-maximization (EM)
have been proposed to overcome
these limitations, but are often computationally demanding
since the E-step in the EM procedure typically
involves state estimation for a high-dimensional nonlinear dynamical
system.  In this work, we propose a computationally fast method for the
state estimation based on  a hybrid of
loopy belief propagation and approximate message passing (AMP).
The key insight is that
a neural system as viewed through calcium imaging can be factorized into simple
scalar dynamical systems for each neuron with linear interconnections
between the neurons.  Using the structure, the updates in
the proposed hybrid AMP methodology can be computed by a set of
one-dimensional state estimation procedures and linear transforms
with the connectivity matrix.  This yields a computationally
scalable method for inferring connectivity of large neural circuits.
Simulations of the method on realistic
neural networks demonstrate good accuracy with computation times
that are potentially
significantly faster than current approaches based on Markov
Chain Monte Carlo methods.
\end{abstract}

\section{Introduction}

Determining connectivity in  populations of neurons
is fundamental to understanding neural computation and function.
In recent years, calcium imaging has emerged as a promising technique for
measuring synaptic activity and mapping neural micro-circuits~\cite{tsien1989fluorescent,ohki2005functional,soriano2008development,vogelstein2009oopsi,stosiek2003vivo}.
Fluorescent calcium-sensitive dyes and genetically-encoded calcium indicators
can be loaded into neurons, which can then be imaged for spiking activity
either \emph{in vivo} or \emph{in vitro}.
Current methods enable imaging populations of hundreds to thousands of neurons
with very high spatial resolution.  Using two-photon microscopy,
imaging can also be localized to specific depths and cortical layers
\cite{svoboda2006principles}.
Calcium imaging also has the potential to
be combined with optogenetic stimulation techniques such as in
\cite{yizhar2011optogenetics}.

However, inferring neural connectivity from calcium imaging
remains a mathematically and computationally
challenging problem.  Unlike anatomical methods,
calcium imaging does not directly measure connections.
Instead, connections must
be inferred indirectly from statistical relationships between spike activities
of different neurons.  In addition, the measurements of the spikes
from calcium imaging are indirect and noisy.
Most importantly, the imaging
introduces significant temporal blurring of the spike times:
the typical time constants for the decay of the fluorescent calcium
concentration, $\Ca2$, can be on the order of a second --
orders of magnitude slower than the spike rates and inter-neuron dynamics.
Moreover, the calcium imaging frame rate remains relatively slow
-- often less than 100~Hz.  Hence, determining connectivity
typically requires super-resolution of spike times within the frame period.

To overcome these challenges, the recent work \cite{MisVogPan:11}
proposed a Bayesian inference method to estimate functional connectivity
from calcium imaging in a systematic manner.
Unlike ``model-free" approaches such as in~\cite{stetter2012model},
the method in \cite{MisVogPan:11}
assumed a detailed functional model of the neural dynamics
with unknown parameters including a connectivity weight matrix $\Wbf$.
The model parameters including the connectivity matrix can
then be estimated via a standard EM procedure~\cite{DempLR:77}.
While the method is general, one of the challenges in implementing the algorithm
is the computational complexity.
As we discuss below, the E-step in the EM procedure
essentially requires estimating the distributions of hidden
states in a nonlinear dynamical
system whose state dimension grows linearly with the number of neurons.
Since exact computation of these densities grows exponentially in the state
dimension, \cite{MisVogPan:11} uses an approximate method based on
blockwise Gibbs sampling where each block of variables
consists of the hidden states associated with one neuron.
Since the variables within a block are described as a low-dimensional dynamical
system, the updates of the densities for the Gibbs sampling
can be computed efficiently via a standard
particle filter \cite{doucet2000sequential,doucet2009tutorial}.
However, simulations of the method show that
the mixing between blocks can still take considerable time to converge.

This paper presents two novel contributions that can potentially
significantly improve the computation time of the EM estimation
as well as the generality of the model.

The first contribution is to employ an approximate message passing (AMP)
technique in the computationally difficult EM step.
The key insight here is to recognize that a system with multiple neurons
can be ``factorized'' into simple, scalar dynamical systems
for each neuron with linear interactions between the neurons.
As described below, we assume a standard leaky
integrate-and-fire (LIF) model for each neuron \cite{DayanAbbott:01} and a
first-order AR process for the calcium imaging~\cite{vogelstein2009spike}.
Under this model, the dynamics of $N$ neurons can be described by $2N$
systems, each with a scalar (i.e.\ one-dimensional) state.  The coupling
between the systems will be linear as described by the connectivity
matrix $\Wbf$.
Using this factorization, approximate state estimation can then be efficiently
performed via approximations of
loopy belief propagation (BP) \cite{WainwrightJ:08}.
Specifically, we show that the loopy BP
updates at each of the factor nodes associated with the
integrate-and-fire and calcium imaging can be performed via a scalar standard
forward--backward filter.  For the updates associated with
the linear transform $\Wbf$,
we use recently-developed approximate message passing (AMP) methods.

AMP was originally proposed in \cite{DonohoMM:09} for problems in compressed
sensing.  Similar to expectation propagation \cite{Minka:01},
AMP methods use Gaussian and quadratic approximations of loopy BP
but with further simplifications that leverage the linear interactions.
AMP was used for neural mapping from multi-neuron excitation and
neural receptive field estimation in~\cite{FletcherRVB:11,KamRanFU:12-nips}.
Here, we use a so-called hybrid AMP technique proposed in \cite{RanganFGS:12-ISIT}
that combines AMP updates across the linear coupling terms with standard
loopy BP updates on the remainder of the system.  When applied to the neural
system, we show that the estimation updates become remarkably simple:
For a system with $N$ neurons, each iteration involves running
$2N$ forward--backward scalar state estimation algorithms,
along with multiplications by $\Wbf$ and $\Wbf^T$ at each time step.
The practical complexity scales as $O(NT)$ where $T$ is the number of time steps.
We demonstrate that the method can be significantly faster than
the blockwise Gibbs sampling proposed in \cite{MisVogPan:11}, with similar accuracy.

In addition to the potential computational improvement,
the AMP-based procedure is somewhat more general.  For example,
the approach in \cite{MisVogPan:11} assumes a generalized linear model (GLM)
for the spike rate of each neuron.  The approach in this work can be theoretically
applied to arbitrary scalar dynamics that describe spiking.  In particular,
the approach can incorporate a physically more realistic LIF model.

The second contribution is a novel method for initial estimation of the connectivity matrix.
Since we are applying the EM methodology to a fundamentally non-convex problem,
the algorithm is sensitive to the initial condition.
However, there are now several good approaches for initial estimation the spike times of each neuron
from its calcium trace via sparse deconvolution
\cite{vogelstein2010fast,onativia2013finite,pnevmatikakis2013sparse}.
We show that, under a leaky integrate and fire model,
that if the true spike times were known exactly,
then the maximum likelihood (ML) estimation
of the connectivity matrix can be performed via sparse probit regression -- a standard
convex programming problem used in classification \cite{Bishop:06}.
We propose to obtain an initial estimate for the connectivity matrix $\Wbf$
by applying the sparse probit regression to the initial estimate of the spike times.

\section{System Model} \label{sec:model}

We consider a recurrent network of $N$ spontaneously firing neurons.
All dynamics are approximated in discrete time
with some time step $\Delta$, with a typical value $\Delta$ = 1 ms.
Importantly, this time step is typically
smaller than the calcium imaging period, so the model captures the dynamics
between observations.
Time bins are indexed by
$k=0,\ldots,T-1$, where $T$ is the number of time bins so that $T\Delta$ is the total observation
time in seconds.  Each neuron $i$ generates a sequence of spikes (action potentials) indicated
by random variables $s_i^k$ taking values $0$ or $1$ to represent whether there was a spike in time bin $k$ or
not.  It is assumed that the discretization step $\Delta$ is sufficiently small such that
there is at most one action potential from a neuron in any one time bin.
The spikes are generated via a standard leaky integrate-and-fire (LIF) model
\cite{DayanAbbott:01} where
the (single compartment) membrane voltage $v_i^k$ of each neuron $i$
and its corresponding spike output sequence $s_i^k$ evolve as
\beq \label{eq:vlif}
    \vtilde_i^{\kp1} = (1-\alpha_{IF})v_i^k + q_i^k + d_{v_i}^k,  \quad
    q_i^k = \sum_{j=1}^N W_{ij}s_j^{k-\delta} + b_{IF,i},
    \quad d_{v_i}^k \sim {\mathcal N}(0,\tau_{IF}),
\eeq
and
\beq  \label{eq:vireset}
    (v_i^{\kp1},s_i^{\kp1}) = \begin{cases}
        (\vtilde_i^k, 0) & \mbox{if } v_i^k < \mu, \\
        (0, 1) & \mbox{if } \vtilde_i^k \geq \mu,\end{cases}
\eeq
where $\alpha_{IF}$ is a time constant for the integration leakage;
$\mu$ is the threshold potential at which the neurons spikes;
$b_{IF,i}$ is a constant bias term; $q_i^k$ is the increase in the membrane
potential from the pre-synaptic spikes from other
neurons and $d_{v_i}^k$ is a noise term including both thermal noise and currents from
other neurons that are outside the observation window.
The voltage has been scaled so that the reset voltage is zero.
The parameter $\delta$ is the integer delay (in units of the time step $\Delta$)
between the spike in one neuron
and the increase in the membrane voltage in the post-synaptic neuron.
An implicit assumption in this model is the post-synaptic current arrives in a single time bin
with a fixed delay.

To determine functional connectivity, the key parameter to estimate will be the matrix $\Wbf$
of the weighting terms $W_{ij}$ in \eqref{eq:vlif}.  Each parameter $W_{ij}$ represents
the increase in
the membrane voltage in neuron $i$ due to the current triggered from a spike in neuron $j$.
The connectivity weight $W_{ij}$ will be zero whenever neuron $j$ has no connection to
neuron $i$.
Thus, determining $\Wbf$ will determine which neurons are connected to one another and the
strengths of those connections.


For the calcium imaging, we use a standard model~\cite{MisVogPan:11},
where the concentration of fluorescent Calcium
has a fast initial rise upon an action potential followed by a slow exponential decay.
Specifically, we let $z_i^k=\Ca2_k$ be the concentration of fluorescent Calcium
in neuron $i$
in time bin $k$ and assume it evolves as first-order auto-regressive $AR(1)$ model,
\beq \label{eq:zk}
    z_i^{\kp1} = (1-\alpha_{CA,i})z_i^k + s_i^k,
\eeq
where $\alpha_{CA}$ is the Calcium time constant.
The observed net fluorescence level is then given by a noisy version of $z_i^k$,
\beq \label{eq:yk}
    y_i^k = a_{CA,i} z_i^k + b_{CA,i} + d_{y_i}^k, \quad d_{y_i}^k \sim {\mathcal N}(0,\tau_y),
\eeq
where $a_{CA,i}$ and $b_{CA,i}$ are constants and $d_{y_i}$ is
white Gaussian noise with variance $\tau_y$.
Nonlinearities such as saturation described in
\cite{vogelstein2009spike} can also be modeled.

As mentioned in the Introduction, a key challenge in calcium imaging is the relatively slow
frame rate which has the effect of subsampling of the fluorescence.  To model the subsampling,
we let $I_F$ denote the set of time indices $k$ on which we observe $F_i^k$.
We will assume that fluorescence values are observed once every $T_F$ time steps for some integer
period $T_F$ so that $I_F = \left\{ 0, T_F, 2T_F, \ldots, KT_F\right\}$
where $K$ is the number of Calcium image frames.

\section{Parameter Estimation via Message Passing}

\subsection{Problem Formulation}

Let $\theta$ be set of all the unknown parameters,
\beq \label{eq:thetaDef}
    \theta = \{ \Wbf, \tau_{IF}, \tau_{CA}, \alpha_{IF}, b_{IF,i},
        \alpha_{CA}, a_{CA,i}, b_{CA,i}, i=1,\ldots,N \},
\eeq
which includes the connectivity matrix, time constants
and various variances and bias terms.
Estimating the parameter set $\theta$ will provide an estimate of the
connectivity matrix $\Wbf$, which is our main goal.

To estimate $\theta$,
we consider a regularized maximum likelihood (ML) estimate
\beq \label{eq:thetaML}
    \thetahat = \argmax_{\theta} L(\ybf|\theta) + \phi(\theta), \quad
    L(\ybf|\theta) = -\log p(\ybf|\theta),
\eeq
where $\ybf$ is the set of observed values;
$L(\ybf|\theta)$ is the negative log likelihood of $\ybf$ given the parameters
$\theta$ and $\phi(\theta)$ is some regularization function.
For the calcium imaging problem, the observations $\ybf$ are the observed
fluorescence values across all the neurons,
\beq \label{eq:yveci}
    \ybf = \left\{ \ybf_1,\ldots,\ybf_N \right\}, \quad
    \ybf_i = \left\{ y_i^k, \quad k \in I_F \right\},
\eeq
where $\ybf_i$ is the set of fluorescence values from neuron $i$, and,
as mentioned above, $I_F$ is the set of time indices $k$ on which the fluorescence
is sampled.

The regularization function $\phi(\theta)$ can be used to impose
constraints or priors on the parameters.  In this work, we will assume a simple
regularizer that only constrains the connectivity matrix $\Wbf$,
\beq \label{eq:phiell1}
    \phi(\theta) = \lambda \|\Wbf\|_1, \quad \|\Wbf\|_1 := \sum_{ij} |W_{ij}|,
\eeq
where $\lambda$ is a positive constant.
The $\ell_1$ regularizer is a standard convex
function used to encourage sparsity \cite{Donoho:06}, which we know
in this case must be valid since most neurons are not connected to one another.

\subsection{EM Estimation}
Exact computation of $\thetahat$ in \eqref{eq:thetaML} is generally intractable, since
the observed fluorescence values $\ybf$ depend on the unknown parameters $\theta$
through a large set of hidden variables.
Similar to \cite{MisVogPan:11}, we thus use a standard EM procedure \cite{DempLR:77}.
To apply the EM procedure to the calcium imaging problem,
let $\xbf$ be the set of hidden variables,
\beq \label{eq:xdef}
    \xbf = \left\{ \vbf, \zbf, \qbf, \sbf \right\},
\eeq
where $\vbf$ are the membrane voltages of the neurons,
$\zbf$ the calcium concentrations, $\sbf$ the  spike outputs and $\qbf$ the linearly
combined spike inputs.  For any of these variables,
we will use the subscript $i$ (e.g.\ $\vbf_i$) to denote the values
of the variables of a particular neuron $i$ across all time steps
and superscript $k$ (e.g.\ $\vbf^k$) to denote the values across all neurons
at a particular time step $k$.  Thus, for the membrane voltage
\[
    \vbf = \left\{v_i^k\right\}, \quad
    \vbf^k = \left(v_1^k, \ldots, v_{N}^k\right), \quad
    \vbf_i = \left( v_i^0,\ldots,v_i^{\Tm1}\right).
\]

The EM procedure alternately estimates distributions on the hidden variables $\xbf$
given the current parameter estimate for $\theta$ (the E-step);
and then updates the estimates for parameter vector $\theta$ given the
current distribution on the hidden variables $\xbf$ (the M-step).
\begin{itemize}
\item \emph{E-Step:}  Given parameter estimates $\thetahat^\ell$, estimate
\beq \label{eq:estepDist}
    P(\xbf|\ybf,\thetahat^\ell),
\eeq
which is the posterior distribution of the hidden variables $\xbf$
given the observations $\ybf$ and current parameter estimate $\thetahat^\ell$.

\item \emph{M-step} Update the parameter estimate via the minimization,
\beq \label{eq:mstep}
    \thetahat^{\ell+1} = \argmin_\theta \Exp\left[ L(\xbf,\ybf|\theta) | \thetahat^{\ell}
        \right] + \phi(\theta),
\eeq
where $L(\xbf,\ybf|\theta)$ is the joint negative log likelihood,
\beq \label{eq:Lxytheta}
    L(\xbf,\ybf|\theta) = - \log p(\xbf,\ybf|\theta).
\eeq
In \eqref{eq:mstep} the expectation is with respect to the distribution
found in \eqref{eq:estepDist} and $\phi(\theta)$ is the parameter
regularization function.
\end{itemize}
The next two sections will describe how we approximately
perform each of these steps.

\subsection{E-Step estimation via Approximate Message Passing}
\label{sec:estep}

For the calcium imaging problem,
the challenging step of the EM procedure is the E-step,
since the hidden variables $\xbf$ to be estimated
are the states and outputs of a high-dimensional nonlinear dynamical system.
Under the model in Section~\ref{sec:model},
a system with $N$ neurons will require $N$ states for the membrane
voltages $v_i^k$ and $N$ states for the bound Ca concentration levels
$z_i^k$, resulting in a total state dimension of $2N$.
The E-step for this system is essentially a state estimation problem,
and exact inference of the states of a general nonlinear dynamical system
grows exponentially in the state dimension.
Hence, exact computation of the posterior distribution \eqref{eq:estepDist}
for the system will be intractable even for a moderately sized network.

As described in the Introduction,
we thus use an approximate messaging passing method that exploits
the separable structure of the system.
For the remainder of this section, we will assume the parameters
$\theta$ in \eqref{eq:thetaDef} are fixed to the current parameter estimate
$\thetahat^\ell$.
Then, under the assumptions of Section~\ref{sec:model},
the joint probability distribution function
of the variables can be written in a factorized form,
\beq \label{eq:pxyfact}
    P(\xbf,\ybf) = P(\qbf,\vbf,\sbf,\zbf,\ybf)
        = \frac{1}{Z} \prod_{k=0}^{\Tm1} \indic{\qbf^k = \Wbf\sbf^k}
    \prod_{i=1}^N \psi^{IF}_i(\qbf_i,\vbf_i,\sbf_i)
        \psi^{CA}_i(\sbf_i,\zbf_i,\ybf_i),
\eeq
where $Z$ is a normalization constant; $\psi^{IF}_i(\qbf_i,\vbf_i,\sbf_i)$
is the potential
function relating the summed spike inputs $\qbf_i$ to the membrane
voltages $\vbf_i$ and spike outputs $\sbf_i$;
$\psi^{CA}_i(\sbf_i,\zbf_i,\ybf_i)$ relates the spike outputs $\sbf_i$
to the bound calcium concentrations $\zbf_i$ and
observed fluorescence values $\ybf_i$; and
the term $\indic{\qbf^k = \Wbf\sbf^k}$ indicates that the distribution
is to be restricted to the set satisfying the linear constraints
$\qbf^k = \Wbf\sbf^k$ across all time steps $k$.

As in standard loopy BP \cite{WainwrightJ:08},
we represent the distribution \eqref{eq:pxyfact} in a \emph{factor graph}
as shown in Fig.~\ref{fig:factorGraph}.
Now, for the E-step, we need to compute the marginals of
the posterior distribution $p(\xbf|\ybf)$ from the joint distribution \eqref{eq:pxyfact}.
Using the factor graph representation, loopy BP iteratively updates
 estimates of these marginal posterior distributions using a message
passing procedure, where the estimates of the distributions
(called beliefs) are passed between the variable and factor nodes
in the graph.

To reduce the computations in loopy BP further, we employ an
approximate message passing (AMP) method for the updates in the factor node
corresponding to the linear constraints $\qbf^k = \Wbf\sbf^k$.
AMP was originally developed in \cite{DonohoMM:09} for problems in compressed sensing,
and can be derived as Gaussian approximations of loopy BP
\cite{Montanari:12-bookChap,Rangan:11-ISIT} similar to expectation propagation \cite{Seeger:08}.
In this work, we employ a hybrid form of AMP \cite{RanganFGS:12-ISIT} that combines AMP with
standard message passing.
The AMP methods have the benefit of being computationally very fast and, for
problems with certain large random transforms, the methods can yield provably Bayes-optimal
estimates of the posteriors, even in certain non-convex problem instances.
However, similar to standard loopy BP, the AMP and its variants may diverge for
general transforms (see \cite{RanSRFC:13-ISIT,Krzakala:14-ISITbethe,RanSchFle:14-ISIT}
for some discussion of the convergence).  For our problem, we will see in simulations
that we obtain fast convergence in a relatively small number of iterations.

\tikzstyle{varNode}=[circle,draw,fill=orange!30]
\tikzstyle{obsVarNode}=[circle,draw,fill=orange!70]
\tikzstyle{factNode}=[rectangle,draw,fill=green!30]
\def\rowShift{0.5cm}
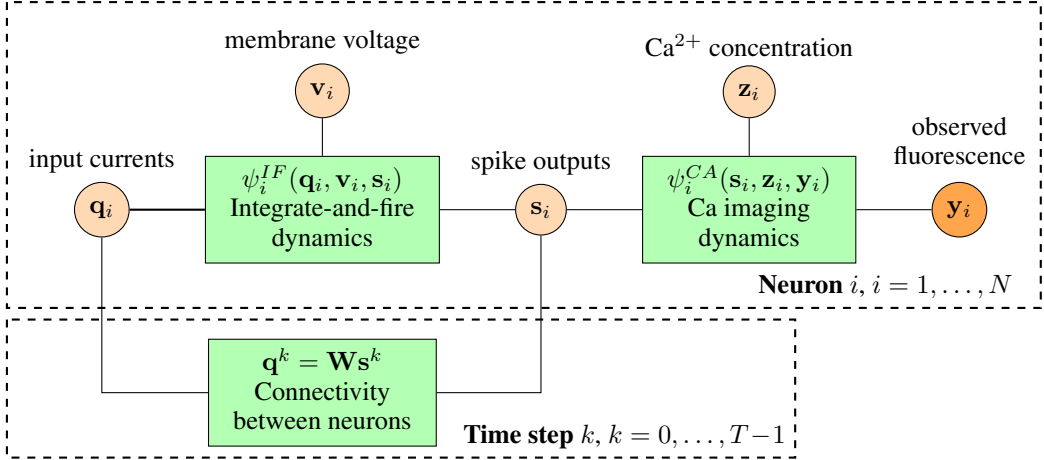
\begin{figure}
\center
\begin{tikzpicture}[scale=1]

\node [varNode,label={[above] input currents}] (qi) {$\qbf_i$};
\node [factNode, right=1cm of qi] (psiIF)
    {\begin{tabular}{c}
    $\psi^{IF}_i(\qbf_i,\vbf_i,\sbf_i)$ \\
    Integrate-and-fire\\ dynamics \end{tabular} };
\node [varNode, above=0.5cm of psiIF,label={[above] membrane voltage}]
    (vi) {$\vbf_i$};
\node [varNode, right=1cm of psiIF,label={[above] spike outputs}]
    (si) {$\sbf_i$};
\node [factNode, below=1cm of psiIF] (W)
    {\begin{tabular}{c} $\qbf^k=\Wbf\sbf^k$ \\ Connectivity \\ between neurons
     \end{tabular} };
\node [factNode, right=1cm of si] (psiCA)  {\begin{tabular}{c}
    $\psi^{CA}_i(\sbf_i,\zbf_i,\ybf_i)$ \\
    Ca imaging\\ dynamics \end{tabular} };
\node [obsVarNode, right=1cm of psiCA,label={[above]
    \begin{tabular}{c} observed \\ fluorescence \end{tabular}}]
    (yi) {$\ybf_i$};
\node [varNode, above=0.5cm of psiCA,label={[above] Ca$^{2+}$ concentration}]
    (zi) {$\zbf_i$};
\node at (psiCA) [below right=0.75cm and 0.0cm] (nlabel)
    {\textbf{Neuron} $i$, $i=1,\ldots,N$};
\node at (W) [below right=0.3cm and 1.75cm] (klabel)
    {\textbf{Time step} $k$, $k=0,\ldots,\Tm1$};

\draw [thick,-] (qi) -- (psiIF);
\draw [-] (vi) -- (psiIF);
\draw [-] (si) -- (psiIF);
\draw [-] (si) |- (W);
\draw [-] (qi) |- (W);
\draw [-] (si) -- (psiCA);
\draw [-] (yi) -- (psiCA);
\draw [-] (zi) -- (psiCA);

\draw [thick,dashed]  ($(qi.north west)+(-1,2.5)$)
    rectangle ($(nlabel.south east)+(0.25,0)$);
\draw [thick,dashed] ($(qi.south west)+(-1,-1.2)$)
    rectangle ($(klabel.south east)$);

\end{tikzpicture}
\caption{Factor graph plate representation of the system
where the spike dynamics are described by the factor node
$\psi^{IF}_i(\qbf_i,\vbf_i,\sbf_i)$ and the calcium image dynamics
are represented via the factor node $\psi^{CA}_i(\sbf_i,\zbf_i,\ybf_i)$.
The high-dimensional dynamical system is described as
$2N$ scalar dynamical systems (2 for each neuron) with linear interconnections,
$\qbf^k=\Wbf\sbf^k$ between the neurons.
A computational efficient
approximation of loopy BP \cite{RanganFGS:12-ISIT} is applied to this
graph for approximate Bayesian inference required in the
E-step of the EM algorithm. }
\label{fig:factorGraph}
\end{figure}

We provide some details of the hybrid AMP method in Appendix~\ref{sec:estepDetails},
but the basic procedure for the factor node updates and the reasons
why these computations are simple can be summarized as follows.
At a high level, the factor graph structure in Fig.~\ref{fig:factorGraph}
partitions the $2N$-dimensional nonlinear dynamical system into
$N$ scalar systems associated with each membrane voltage $v_i^k$
and an additional $N$ scalar systems associated with each calcium
concentration level $z_i^k$.  The only coupling between these
systems is through the linear relationships $\qbf^k=\Wbf\sbf^k$.
As shown in Appendix~\ref{sec:estepDetails},
on each of the scalar systems, the factor node updates
required by loopy BP essentially reduces
to a state estimation problem for this system.  Since the state space
of this system is scalar (i.e. one-dimensional),
we can discretize the state space well with a small number of points
-- in the experiments below we use $L=20$ points per dimension.
Once discretized, the state estimation can be performed via
a standard forward--backward algorithm.  If there are $T$ time steps,
the algorithm will have a computational cost of $O(TL^2)$ per scalar system.
Hence, all the factor node updates across all the $2N$ scalar systems
has total complexity $O(NTL^2)$.

For the factor nodes associated with the linear constraints $\qbf^k=\Wbf\sbf^k$,
we use the AMP approximations \cite{RanganFGS:12-ISIT}.  In this
approximation, the messages for the transform outputs $q_i^k$
are approximated as Gaussians which is, at least heuristically, justified
since the they are outputs of a linear transform of a large number of
variables, $s^k_i$.  In the AMP algorithm, the belief updates for the variables
$\qbf^k$ and $\sbf^k$ can then be computed simply by linear transformations
of $\Wbf$ and $\Wbf^T$.  Since $\Wbf$ represents a connectivity matrix,
it is generally sparse.  If each row of $\Wbf$ has $d$ non-zero values,
multiplication by $\Wbf$ and $\Wbf^T$ will be $O(Nd)$.  Performing the multiplications across all time steps results in a total complexity of $O(NTd)$.

Thus, the total complexity of the proposed E-step estimation method
is $O(NTL^2 + NTd)$ per loopy BP iteration.
We typically use a small number of loopy BP iterations per EM update
(in fact, in the experiments below, we found reasonable performance with
one loopy BP update per EM update).  In summary, we see that while the
overall neural system is high-dimensional, it has a linear + scalar structure.
Under the assumption of the bounded connectivity $d$,
this structure enables an approximate
inference strategy that scales linearly with the number of neurons $N$
and time steps $T$.  Moreover, the updates in different scalar systems
can be computed separately allowing a readily parallelizable implementation.

\subsection{Approximate M-step Optimization} \label{sec:mstep}
The M-step \eqref{eq:mstep} is computationally relatively simple.
All the parameters in $\theta$ in \eqref{eq:thetaDef} have a
linear relationship between the components of the variables in the
vector $\xbf$ in \eqref{eq:xdef}.  For example, the parameters
$a_{CA,i}$ and $b_{CA,i}$ appear in the fluorescence output
equation \eqref{eq:yk}.  Since the noise
$d_{y_i}^k$ in this equation is Gaussian, the negative
log likelihood \eqref{eq:Lxytheta} is given by
\[
    L(\xbf,\ybf|\theta) = \frac{1}{2\tau_{y_i}} \sum_{k \in I_F}
        (y_i^k - a_{CA,i}z_i^k-b_{CA,i})^2 + \frac{T}{2}\log(\tau_{y_i}) +
        \mbox{other terms},
\]
where ``other terms" depend on parameters other than $a_{CA,i}$ and
$b_{CA,i}$.
The expectation $\Exp( L(\xbf,\ybf|\theta)|\thetahat^\ell )$
will then depend only on the mean and variance of the variables $y_i^k$
and $z_i^k$, which are provided by the E-step estimation.
Thus, the M-step optimization in \eqref{eq:mstep} can be computed via
a simple least-squares problem.  Using the linear relation
\eqref{eq:vlif}, a similar method
can be used for $\alpha_{IF,i}$ and $b_{IF,i}$, and the linear relation
\eqref{eq:zk} can be used to estimate the calcium time constant $\alpha_{CA}$.

To estimate the connectivity matrix $\Wbf$, let $\rbf^k = \qbf^k-\Wbf\sbf^k$
so that the constraints in \eqref{eq:pxyfact} is equivalent to the
condition that $\rbf^k=0$.  Thus, the term containing $\Wbf$ in the
expectation of the negative log likelihood
$\Exp( L(\xbf,\ybf|\theta)|\thetahat^\ell )$ is given by the
negative log probability density of $\rbf^k$ evaluated at zero.
In general, this density will be a complex function of $\Wbf$ and difficult
to minimize.  So, we approximate the density as follows:  Let $\qbfhat$
and $\sbfhat$ be the expectation of the variables $\qbf$ and $\sbf$ given by
the E-step.  Hence, the expectation of $\rbf^k$ is $\qbfhat^k-\Wbf\sbfhat^k$.
As a simple approximation, we will then assume that the variables $r_i^k$
are Gaussian, independent and having some constant variance $\sigma^2$.
Under this simplifying assumption, the M-step optimization of $\Wbf$
with the $\ell_1$ regularizer \eqref{eq:phiell1} reduces to
\beq \label{eq:What}
    \widehat{\Wbf} = \argmin_{\Wbf} \frac{1}{2}\sum_{k=0}^{\Tm1}
    \|\qbfhat^k - \Wbf\sbfhat^k\|^2
    + \sigma^2\lambda \|\Wbf\|_1,
\eeq
For a given value of $\sigma^2\lambda$, the optimization \eqref{eq:What}
is a standard LASSO optimization \cite{Tibshirani:96} which can be evaluated
efficiently via a number of convex programming methods.  In this
work, in each M-step, we adjust the regularization parameter $\sigma^2\lambda$
to obtain a desired fixed sparsity level in the solution $\Wbf$.

\subsection{Initial Estimation via Sparse Regression} \label{sec:initW}
Since the EM algorithm cannot be guaranteed to converge a global maxima,
it is important to pick the initial parameter estimates carefully.
The time constants and noise levels for the calcium image can be extracted
from the second-order statistics of fluorescence values and simple thresholding
can provide a coarse estimate of the spike rate.

The key challenge is to obtain a good estimate for the connectivity
matrix $\Wbf$.
For each neuron $i$, we first make an initial estimate
of the spike probabilities $P(s_i^k=1|\ybf_i)$
from the observed fluorescence values $\ybf_i$,
assuming some i.i.d.\ prior of the form
$P(s_i^t)=\lambda\Delta$, where $\lambda$
is the estimated average spike rate per second.
This estimation can be solved with
the filtering method in \cite{vogelstein2009spike} and is also equivalent to
the method we use for the factor node updates.
We can then threshold these probabilities to make a hard initial decision
on each spike:  $s_i^k=0$ or 1.

We then propose to estimate $\Wbf$
from the spikes as follows.
Fix a neuron $i$ and let $\wbf_i$ be the vector
of weights $W_{ij}$, $j=1,\ldots,N$.  Under the assumption that the initial spike
sequence $s_i^k$ is exactly correct, it is shown in
Appendix~\ref{sec:initWApp} that a regularized maximum likelihood estimate of $\wbf_i$
and bias term $b_{IF,i}$ is given by
\beq \label{eq:betaOpt1}
    (\wbfhat_i,\bhat_{IF,i}) = \argmin_{\wbf_i,b_{IF,i}}
        \sum_{k=0}^{\Tm1} L_{ik}( \ubf_{k}^T \wbf_i +c_{ik}b_{IF,i}  - \mu, s_i^k)
    + \lambda\sum_{j=1}^N |W_{ij}|,
\eeq
where $L_{ik}$ is a probit loss function and the vector $\ubf_k$
and scalar $c_{ik}$ can be determined from the spike estimates.
The optimization \eqref{eq:betaOpt1} is precisely a standard
probit regression used in sparse linear classification \cite{Bishop:06}.
This form arises due to the nature of the leaky integrate-and-fire model
\eqref{eq:vlif} and \eqref{eq:vireset}.
Thus, assuming the initial spike sequences are estimated reasonably
accurately, one can obtain good initial estimates for the weights $W_{ij}$ and
bias terms $b_{IF,i}$ by solving a standard classification problem.

We point out that \cite{SoltaniGold:14} has recently provided an alternative method for
recovery of connectivity matrix from the spikes assuming a LIF model
based on maximizing information flow.

\begin{table}
\begin{center}
\small
\begin{tabular}{|p{2in}|p{3.1in}|}
\hline
\textbf{Parameter} & \textbf{Value} \\ \hline
Number of neurons, $N$ & 100 \\ \hline
Connection sparsity & 10\% with random connections. All connections
    are excitatory with the non-zero weights $W_{ij}$ being
    exponentially distributed.   \\ \hline
Mean firing rate per neuron & 10 Hz \\ \hline
Simulation time step, $\Delta$ & 1 ms \\ \hline
Total simulation time, $T\Delta$ & 10 sec (10,000 time steps) \\ \hline
Integration time constant, $\alpha_{IF}$ & 20 ms \\ \hline
Conduction delay, $\delta$ & 2 time steps = 2 ms \\ \hline
Integration noise, $d_{v_i}^k$ & Produced from two unobserved neurons. \\ \hline
Ca time constant, $\alpha_{CA}$ & 500 ms \\ \hline
Fluorescence noise, $\tau_{CA}$ & Set to 20 dB SNR \\ \hline
Ca frame rate , $1/T_F$ & 100 Hz \\ \hline
\end{tabular}
\end{center}
\caption{Parameters for the Ca image simulation. \label{tbl:simParam}}
\end{table}

\begin{SCfigure}
	\includegraphics[width=0.6\textwidth]{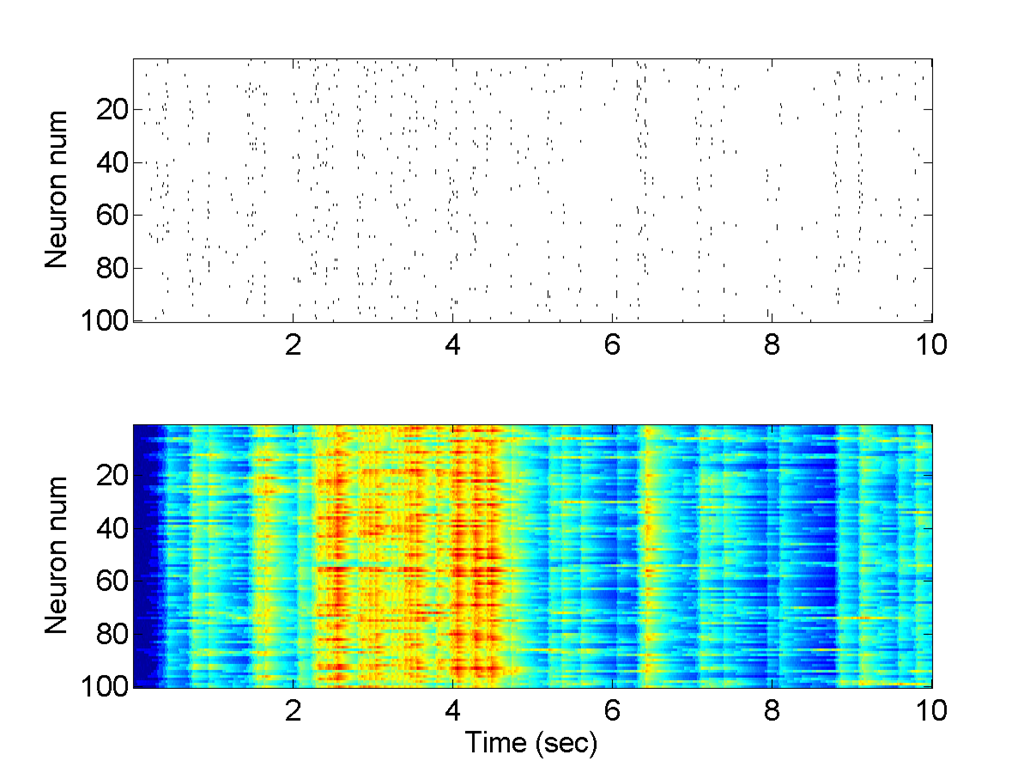}
 \caption{Typical network simulation trace.
    Top panel: Spike traces for the 100 neuron simulated network.
    Bottom panel:  Calcium image fluorescence levels.
    Due to the random network topology, neurons often fire together,
    significantly complicating connectivity detection.  Also,
    as seen in the lower panel, the slow decay of the fluorescent calcium
    blurs the spikes in the calcium image.  \vspace{1cm} \label{fig:CaNetSim}
 }
\end{SCfigure}

\section{Numerical Example}

The method was tested using realistic network parameters, as shown in
Table~\ref{tbl:simParam},
similar to those found in neurons networks within a cortical
column~\cite{sayer1990time}.  Similar parameters are used in \cite{MisVogPan:11}.
The network consisted of 100 neurons with each neuron randomly connected
to 10\% of the other neurons.  The non-zero weights $W_{ij}$ were drawn
from an exponential distribution.  All weights
were positive (i.e. the neurons were excitatory -- there were no inhibitory
neurons in the simulation).  However, inhibitory neurons can also be added.
A typical random matrix $\Wbf$ generated in this
manner would not in general result in a stable system.  To stabilize the system,
we followed the procedure in \cite{stetter2012model} where the
system is simulated multiple times.  After each simulation,
the rows of the matrix $\Wbf$ were adjusted up or down to increase or decrease
the spike rate until all neurons spiked at a desired target rate.
In this case, we assumed a desired average spike rate of 10~Hz.

From the parameters in Table~\ref{tbl:simParam}, we can immediately see
the challenges in the estimation.  Most importantly, the calcium imaging
time constant $\alpha_{CA}$ is set for 500~ms.  Since the average
neurons spike rate is assumed to be 10~Hz,
several spikes will typically appear within a single time constant.
Moreover, both the integration time constant and inter-neuron conduction time
are much smaller than both the image frame rate and Calcium time constants.

A typical simulation of the network after the stabilization
is shown in Fig.~\ref{fig:CaNetSim}.
Observe that due to the random connectivity,
spiking in one neuron can rapidly cause the entire network to fire.
This appears as the vertical bright stripes in the lower panel of Fig.~\ref{fig:CaNetSim}.
This synchronization makes the connectivity detection difficult to detect under temporal blurring of Ca imaging
since it is hard to determine which neuron is causing which neuron to fire.
Thus, the random matrix is a particularly challenging test case.

\begin{figure}
\centering
\includegraphics[width=0.4\textwidth]{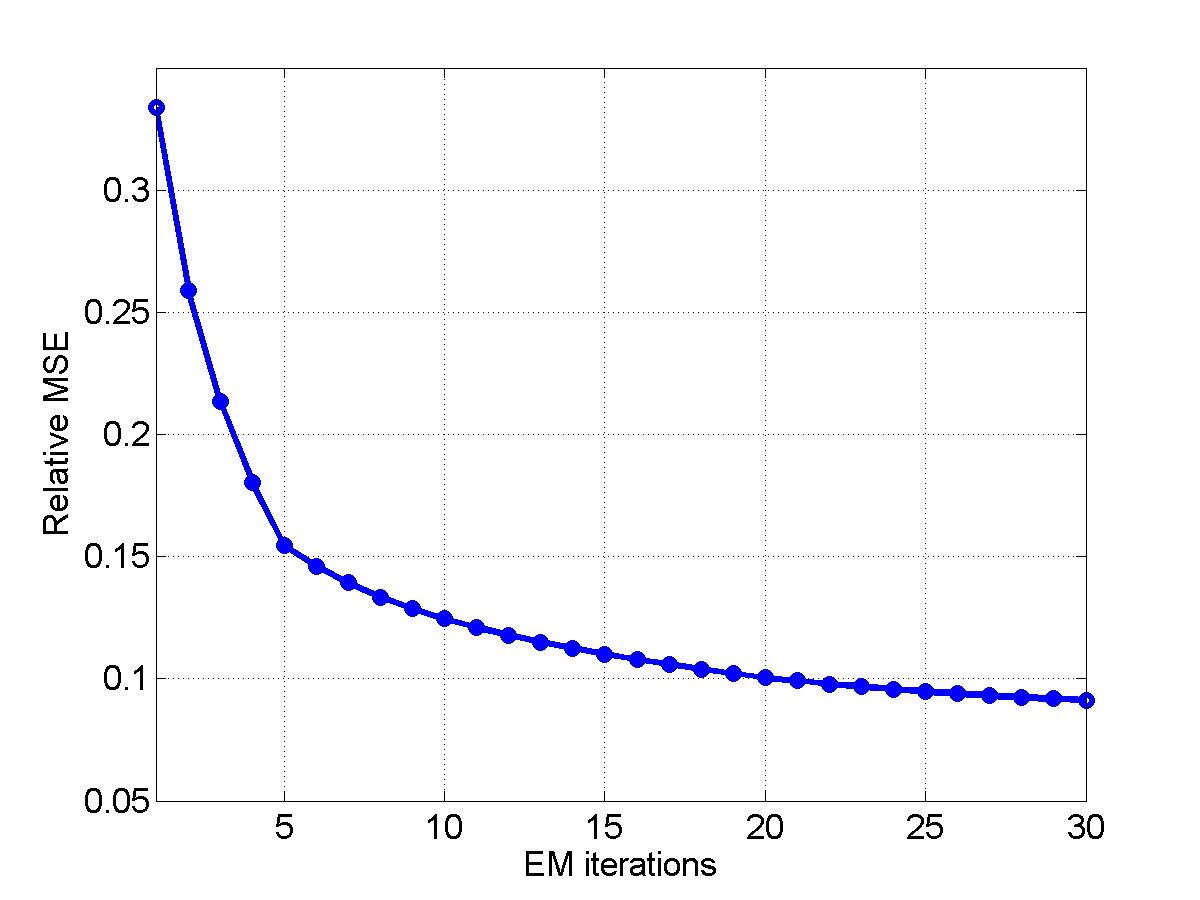}
\includegraphics[width=0.4\textwidth]{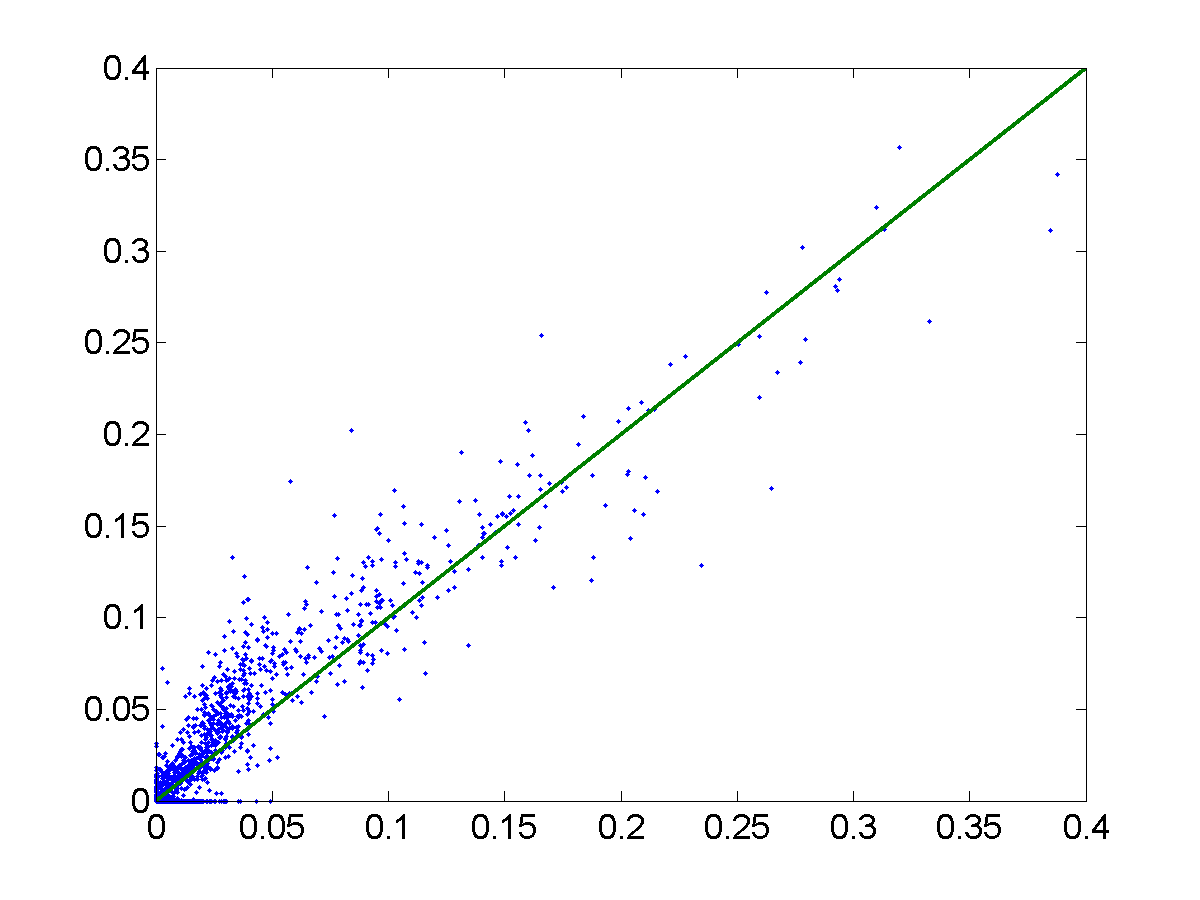}
\caption{Weight estimation accuracy.  Left:  Normalized mean-squared
error as a function of the iteration number.  Right:  Scatter plot
of the true and estimated weights.
\label{fig:simResults} }
\end{figure}

The results of the estimation are shown in Fig.~\ref{fig:simResults}.
The left panel shows the relative mean squared error defined as
\beq \label{eq:relMse}
    \mbox{relative MSE} =
        \frac{\min_{\alpha} \sum_{ij} |W_{ij}-\alpha \What_{ij}|^2 }{
            \sum_{ij} |W_{ij}|^2 },
\eeq
where $\What_{ij}$ is the estimate for the weight $W_{ij}$.
The minimization over all $\alpha$ is performed since the method
can only estimate the weights up to a constant scaling.
The relative MSE is plotted as a function of the EM iteration,
where we have performed only a single loopy BP iteration for each
EM iteration.  We see that after only 30 iterations we obtain a relative
MSE of 7\% -- a number at least comparable to earlier
results in \cite{MisVogPan:11}, but with significantly less computation.
The right panel shows a scatter plot of the estimated weights $\What_{ij}$
against the true weights $W_{ij}$.


\section{Conclusions} We have presented a scalable method
for inferring connectivity in neural systems from calcium imaging.
The method is based on factorizing the systems into scalar dynamical systems
with linear connections.  Once in this form, state estimation -- the key
computationally challenging component of the EM estimation -- is tractable
via approximating message passing methods.
The key next step in the work is to test the methods on real data
and also provide more comprehensive computational comparisons against
current techniques such as \cite{MisVogPan:11}.

\appendix
\section{E-Step Message Passing Implementation Details} \label{sec:estepDetails}

As described in Section~\ref{sec:estep},
the E-step inference is performed via an
approximate message passing technique \cite{RanganFGS:12-ISIT}.
As in standard sum-product loopy BP \cite{WainwrightJ:08},
the algorithm is based on passing ``belief messages" between
the variable and factor nodes representing estimates of the posterior
marginals of the variables.
Referring to the factor graph in Fig.~\ref{fig:factorGraph},
we will use the subscripts $IF$, $CA$ and $W$ to refer respectively
to the factor nodes for integrate and fire potential functions $\psi^{IF}_i$,
the calcium imaging potential functions $\psi^{CA}_i$ and the
linear constraints $\qbf^k=\Wbf\sbf^k$.  We use the subscripts $Q$ and $S$
to refer to the variable nodes for $\qbf$ and $\sbf$.
We use the notation such as $P_{IF \ra Q}(q_i^k)$ to denote the belief
message to the variable
node $q_i^k$ from the integrate and factor node $\psi^{IF}_i$.
Similarly, $P_{IF \la Q}(q_i^k)$ will denote the reverse message from
the variable node to the factor node.

The messages to and from the variable nodes $s_i^k$ are binary:  $s_i^k=0$ or 1.
Hence, they can be parameterized by a single scalar.
Similar to expectation propagation \cite{Minka:01},
the messages to and from the variable nodes $q_i^k$ are approximated as
Gaussians, so that we only need to maintain the first and second moments.
Gaussian approximations are used in the variational Bayes method for calcium
imaging inference in \cite{keshri2013shotgun}.

To apply the hybrid AMP algorithm of \cite{RanganFGS:12-ISIT} to the factor
graph in Fig.~\ref{fig:factorGraph}, we use standard loopy BP message updates
on the IF and CA factor nodes, and AMP updates on the linear constraints
$\qbf^k=\Wbf\sbf^k$.  The AMP updates are based on linear-Gaussian approximations. The details of the messages updates are as follows.

\paragraph{Messages from $\psi^{IF}_i$:}
This factor node represents the integrate and fire system
for the voltages $v_i^k$ and is given by
\beq \label{eq:psiIF}
    \psi^{IF}_i(\qbf_i,\vbf_i,\sbf_i) =
        \prod_{k=0}^{\Tm1} P(v_i^{\kp1},s_i^{\kp1}|v_i^k,q_i^k),
\eeq
where the conditional density $P(v_i^{\kp1},s_i^{\kp1}|v_i^k,q_i^k)$
is given by integrate and fire system \eqref{eq:vlif} and \eqref{eq:vireset}.
To describe the output belief propagation messages for this factor node,
define the joint distribution,
\beqa
    \lefteqn{ P(\qbf_i,\vbf_i,\sbf_i) \propto \psi^{IF}_i(\qbf_i,\vbf_i,\sbf_i)
    P_{IF \la Q}(q_i^k)P_{IF \la S}(s_i^k) } \nonumber \\
    &=& \prod_{k=0}^{\Tm1} P(v_i^{\kp1},s_i^{\kp1}|v_i^k,q_i^k)
        P_{IF \la Q}(q_i^k)P_{IF \la S}(s_i^k),
        \label{eq:pifjoint}
\eeqa
where $P_{IF \la Q}(q_i^k)$ and $P_{IF \la S}(s_i^k)$
are the incoming messages from the variable nodes.
To compute the output messages, we must first compute the
marginal densities $P(q_i^k)$ and $P(s_i^k)$ of this joint
distribution \eqref{eq:pifjoint}.

To compute these marginal densities, define $\xi_i^k = q_i^k + d_{v_i}^k + b_i$.
Now recall that the AMP assumption is that each incoming distribution
$P_{IF \la Q}(q_i^k)$ is Gaussian.  Let $\qhat_i^k$ and $\tau_{q_i}^k$
be the mean and variance of this distribution.
Thus, the joint distribution \eqref{eq:pifjoint} is identical to the posterior
distribution of a linear system with a Gaussian input
\beq \label{eq:vlifi}
    \vtilde_i^k = (1-\alpha_{IF})\vtilde_i^k + \xi_i^k,  \quad
    \xi_i^k \sim {\mathcal N}(\qhat_i^k+b_i,\tau_{q_i}^k + \tau_{IF}),
\eeq
with the reset and spike output in \eqref{eq:vireset}
and output observations $P(s_i^k|\phi_i^k)$.
This is a nonlinear system with a one-dimensional state $v_i^k$.
Hence, one can, in principle, approximately compute the
marginal densities $P(q_i^k)$ and $P(s_i^k)$ of \eqref{eq:pifjoint}
with a one-dimensional particle filter~\cite{doucet2009tutorial}.
However, we found it computationally
faster to simply use a fixed discretization of the set of values $v_i^k$.
In the experiments below we used $L=$ 20 values linearly spaced from 0 to
the threshold level $\mu$.  Using the fixed discretization enables a number
of the computations to be computed once for all time steps,
and also removes the computations and logic for pruning necessary in particle
filtering.  After computing the marginals $P(q_i^k)$ and $P(s_i^k)$,
we set the output messages as
\[
    P_{IF \ra Q}(q_i^k) \propto P(q_i^k)/P_{IF \la Q}(q_i^k), \quad
    P_{IF \ra S}(s_i^k) \propto P(s_i^k)/P_{IF \la S}(s_i^k).
\]

\paragraph{Messages from $\psi^{CA}_i$:}
In this case, the factor node
represents the Ca imaging dynamics and is given by,
\beq \label{eq:psiCA}
    \psi^{CA}_i(\sbf_i,\zbf_i,\ybf_i) =
         \prod_{k=0}^{\Tm1} P(z_i^{\kp1}|z_i^k,s_i^k)
         \prod_{k \in I_F} P(y_i^k|z_i^k),
\eeq
where $P(z_i^{\kp1}|z_i^k,s_i^k)$ and $P(y_i^k|z_i^k)$ are given by
the relations \eqref{eq:zk} and \eqref{eq:yk} describing the
fluorescent Ca$^{2+}$  concentration evolution and observed fluorescence.
Recall that $I_F$ in \eqref{eq:psiCA} is the set of time samples
on which the output $y_i^k$ is observed.
To compute the output beliefs for the factor node, as before, we
define the joint distribution,
\beqa
   \lefteqn{
    P(\sbf_i,\zbf_i,\ybf_i) \propto
       \psi^{CA}_i(\sbf_i,\zbf_i,\ybf_i)\prod_{k=0}^{\Tm1}
       P_{CA \la S}(s_i^k) } \nonumber \\
      &=&  \prod_{k=0}^{\Tm1} P(z_i^{\kp1}|z_i^k,s_i^k)P_{CA \la S}(s_i^k)
       \prod_{k \in I_F} P(y_i^k|z_i^k), \label{eq:pcajoint}
\eeqa
where $P_{CA \la S}(s_i^k)$ are the input messages from the variable nodes
$s_i^k$.
This distribution $P(\sbf_i,\zbf_i,\ybf_i)$ is identical to a
the distribution for a linear system with a scalar state $z_i^k$,
Gaussian observations $y_i^k$ and a discrete zero-one input $s_i^k$
with prior $P_{CA \la S}(s_i^k)$.  Similar to the integrate-and-fire case,
we can approximately compute the posterior marginals $P(s_i^k|\ybf_i)$
by discretizing the states $z_i^k$ and using a standard forward--backward
estimator. From the posterior marginals $P(s_i^k|\ybf_i)$, we can then
compute the belief messages for the factor node back to the variable nodes
$s_i^k$:  $P_{CA \ra S}(s_i^k) \propto P(s_i^k|\ybf_i)/P_{CA \la S}(s_i^k)$.

\paragraph{AMP messages from the linear constraints $\qbf^k=\Wbf\sbf^k$:}
Standard loopy BP updates for this factor node would be intractable
for typical connectivity matrices $\Wbf$.  To see this,
suppose that in the current
estimate for the connection matrix $\Wbf$, each neuron is connected to $d$
other neurons.  Hence the rows of $\Wbf$ will have $d$ non-zero entries.
Each constraints $q_i^k = (\Wbf\sbf^k)_i$ will thus involve $d$ binary variables,
and the complexity of the loopy BP update will then require $O(2^d)$ operations.
This computation will be difficult for large $d$.

The hybrid AMP algorithm of \cite{RanganFGS:12-ISIT} uses Gaussian approximations
on the messages to reduce the computations to simple linear transforms.
First consider the output messages $P_{W \ra Q}(q_i^k)$ to the
variable nodes $q_i^k$.  These messages are Gaussians.  Let $\qhat_i^k$
and $\tau_{q_i^k}$ be their mean and variance and let $\qbfhat^k$ and $\tau_q^k$
be the vector of these quantities.  In the hybrid AMP algorithm, these
means and variances are given by
\beq \label{eq:qgamp}
    \qbfhat^k = \Wbf\sbfhat^k - \taubf_q^k\pbf^k, \quad
    \taubf_q^k = |\Wbf|^2\taubf_s^k,
\eeq
where $\sbfhat^k$ and $\taubf_s^k$ are the vectors of
means and variances from the incoming messages $P_{W \la S}(s_i^k)$,
and $|\Wbf|^2$ is the matrix with components $|W_{ij}|^2$.  The variables
$\pbf^k$ is a real-valued state vector, which is initialized to zero.
In \eqref{eq:qgamp}, the multiplication $\taubf_q^k\pbf^k$
is to be performed componentwise:  $\taubf_q^k\pbf^k)_i = \tau_{q_i^k}p^k_i$.

To process the incoming belief messages from the variable nodes $\qbf^k$,
let $\overline{\qbf}^k$ and $\overline{\taubf}_q^k$ be the
vector of mean and variances of the incoming beliefs
$P_{W \la Q}(q^k_i)$.  These quantities are to be distinguished from
$\qbfhat^k$ and $\tau_q^k$, the mean and variance vectors of the outgoing
messages $P_{W \ra Q}(q^k_i)$.  We then first compute,
\beq
    \pbf^k = (\overline{\qbf}^k - \qbfhat^k)/\taubf_q^k, \quad
    \taubf_p^k = \frac{1}{\taubf_q^k}\left[ 1 -
        \frac{\overline{\taubf}^k_q}{\taubf_q^k} \right],
\eeq
where the divisions are componentwise.
Next, we compute the quantities
\beq
    \overline{\sbf}^k = \sbf^k + \taubf_s^k \Wbf^T\pbf^k, \quad
    \overline{\taubf}_s^k = 1/(|\Wbf|^2\taubf_p^k),
\eeq
where, again, the divisions are componentwise and the multiplication
between $\taubf_s^k$ and $\Wbf^T\pbf^k$ is componentwise.
The output message to the variable nodes $s^k_i$ is then given by
\[
    P_{W \ra S}(s^k_i) \propto \exp\left(
        -\frac{1}{2\taubar_{s_i^k}}(s^k_i - \overline{s}^k_i)^2 \right),
\]
with possible values $s^k_i = 0$ or 1.

\paragraph{Variable node updates:}  The variable node updates
are based on the standard sum-product rule \cite{WainwrightJ:08}.
In the factor graph in Fig.~\ref{fig:factorGraph}, each variable
nodes $q_i^k$ is only connected to two factor nodes:
the factor node for the potential function $\psi^{IF}_i$ and
the factor node for the linear constraint $\qbf^k = \Wbf\sbf^k$.
Hence, the variable node will simply relay the messages between the nodes:
\[
    P_{IF \la Q}(q_i^k) = P_{W \ra Q}(q^k_i), \quad
    P_{IF \la W}(q_i^k) = P_{W \ra IF}(q^k_i),
\]
Recall that these messages are approximated as Gaussians, so the messages
can be represented by mean and variances.

Each binary spike variable nodes $s_i^k$ is connected to three factor nodes:
the integrate and fire potential function $\psi^{IF}_i$,
the calcium imaging potential function $\psi^{CA}_i$ and the
linear constraint $\qbf^k = \Wbf\sbf^k$.  In the sum-product rule,
the output message to any one of these nodes is the product of the incoming
messages from the other two.  Hence,
\beqan
    && P_{IF \la S}(s_i^k) \propto P_{W \ra S}(s_i^k)P_{Q \ra S}(s_i^k), \quad
    P_{CA \la S}(s_i^k) \propto P_{W \ra S}(s_i^k)P_{IF \ra S}(s_i^k), \\
    && P_{W \la S}(s_i^k) \propto P_{CA \ra S}(s_i^k)P_{IF \ra S}(s_i^k).
\eeqan
The proportionality constant is simple to compute since the variables
are binary so that $s_i^k=0$ or 1.

\section{Initial Estimation of $\Wbf$ via Sparse Probit Regression}
\label{sec:initWApp}

We show that given the spike sequence $s_i^k$, the
maximum likelihood estimate of the
connectivity weights $\Wbf$ and bias terms $b_{IF,i}$ can be computed
approximately via a sparse probit regression of the form \eqref{eq:betaOpt1}.
To this end, suppose that we know the true spike sequence $s_i^k$
for all neurons $i$ and times $k$.
Let $\{t_i^{\ell}, \ell=1,\ldots,L_i\}$,
be the index of time bins $k$ where there is a spike (i.e.\ $s_i^k=1$
when $k=t_i^{\ell}$ for some $\ell$).
Now, consider any time $k$ between two spikes $k \in [t_i^\ell,t_i^{\ell+1})$.
Since $s_i^k=1$ at the initial time $k=t_i^\ell$,
\eqref{eq:vireset} shows that the voltage must starts at zero:  $v_i^k=0$.
Integrating \eqref{eq:vlif} from this initial condition, we have that
for any $k \in (t_i^\ell,t_i^{\ell+1})$,
\beq \label{eq:vtildeu}
    \tilde{v}_i^k = \sum_{j=1}^N W_{ij}u_j^k + (k-t_i^\ell)b_{IF,i} + \xi_i^k,
    \quad u_j^k = \sum_{m=0}^{k-t_i^\ell-1} (1-\alpha_{IF})^m s_i^{k-m-\delta},
\eeq
where $\xi_i^k$ is the integration of the Gaussian noise $d_{v_i}^k$ up to
time $k$.
We can rewrite \eqref{eq:vtildeu} in vector form
\beq \label{eq:vtildeuvec}
    \tilde{v}_i^k = \ubf_{k}^T \wbf_i + c_{ik}b_{IF,i} + \xi_i^k,
\eeq
where $\ubf_{k}$ and $\wbf_i$ are the vectors with
the components $W_{ij}$ and $u_j^k$ and $c_{ik} = k-t_i^\ell$.

Now, let $\mathcal A^k$ be the set of spikes $s_j^m$ for all $j$
and all time bins $m \leq k$, so that  ${\mathcal A}_k$
represents the past spike events. Observe that in the model
\eqref{eq:vtildeuvec}, the vector $\ubf_k$
can be computed from ${\mathcal A}_k$ and the noise $\xi^k_i$ is independent
of $\mathcal A^k$.
Also, from \eqref{eq:vireset}, $s_i^{\kp1}=1$ if and only
if $\tilde{v}_i^k \geq \mu$.  Hence, we have that the conditional probability
of the spike event at some time $\kp1$, given the past spikes is
\beq \label{eq:Psreg}
    P(s_i^{\kp1}=1|{\mathcal A}_k) = \Phi\left(
    \frac{\ubf_{k}^T \wbf_i + c_{ik}b_{IF,i} - \mu}{\sigma_{ik}} \right),
\eeq
where $\sigma^2_{ik}$ is the variance of $\xi_i^k$ in \eqref{eq:vtildeuvec},
and $\Phi(z)$ is the cumulative distribution function of a unit Gaussian.
Given the conditional probability \eqref{eq:Psreg},
we can then estimate the parameters $\betabf$, through the maximization
\beq \label{eq:betaOpt}
    (\wbfhat_i,\bhat_{IF,i}) = \argmin_{\wbf_i,b_{IF,i}}
        \sum_{k=0}^{\Tm1} L_{ik}( \ubf_{k}^T \wbf_i +c_{ik}b_{IF,i}  - \mu, s_i^k)
    + \lambda\sum_{j=1}^N |W_{ij}|,
\eeq
where $L_{ij}(z,s)$ is the probit loss function
\beq \label{eq:probitLoss}
    L_{ik}(z,s) = \begin{cases}
        -\log( \Phi(z/\sigma_{ik})) & s_i^k=1 \\
        -\log(1-\Phi(z/\sigma_{ik})) & s_i^k = 0
    \end{cases}
\eeq
Given the conditional probabilities \eqref{eq:Psreg},
the minimization \eqref{eq:betaOpt} is precisely the maximum likelihood
estimate of the parameters with an additional $\ell_1$ regularization
term to encourage sparsity in the weights $\wbf_i$.
But this minimization is exactly a sparse probit regression that is standard in
linear classification~\cite{Bishop:06}.

The only issue is that the optimization function \eqref{eq:betaOpt}
with the probit loss \eqref{eq:probitLoss} requires knowledge of the
threshold $\mu$ and variances $\sigma^2_{ik}$.
Since we are only interested in the connectivity weights up to a constant
factor, we can arbitrarily set the threshold level $\mu$ to some value,
say $\mu=1$.  In principle, the noise variances $\sigma^2_{ik}$
can be derived from the integration noise variance $\tau_{IF}$ in \eqref{eq:vlif}.
However, the variance $\tau_{IF}$ may itself not be initially known.
Instead, we simply select $\sigma^2_{ik}$ to be a constant value that is
relatively large to account for initial errors in the $s_i^k$.

\section*{Acknowledgments} 

This research was supported by NSF grants 1116589 and 1254204. The authors
would like to thank Bruno Olshausen, Fritz Sommer,
Lav Varshney, Mitya Chlovskii, Peyman Milanfar,
Evan Lyall, and Eftychios Pnevmatikakis for their insights and support.
This work would not have been possible without the supportive environment and wonderful discussions at the Berkeley Redwood Center for Theoretical Neuroscience -- thank you.

\small{
\bibliographystyle{IEEEtran}
\bibliography{bibl}
}

\end{document}